\documentclass[12pt,a4paper]{article}

\usepackage[utf8]{inputenc}
\usepackage[T1]{fontenc}
\usepackage{mathptmx}
\usepackage{geometry}
\usepackage{setspace}
\usepackage{graphicx}
\usepackage{booktabs}
\usepackage{hyperref}
\usepackage{natbib}
\setcitestyle{authoryear,round}
\usepackage{xcolor}

\title{Navigating Algorithmic Opacity: Folk Theories and User Agency in Semi-Autonomous Vehicles}

\author{Yehuda Perry, Tawfiq Ammari - Rutgers University}

\date{}

\begin{document}

\maketitle

\begin{abstract}
As semi-autonomous vehicles (AVs) become prevalent, drivers must collaborate with AI systems whose decision-making processes remain opaque. This study examines how drivers of AVs develop folk theories to interpret algorithmic behavior that contradicts their expectations. Through 16 semi-structured interviews with drivers in the United States, we investigate the explanatory frameworks drivers construct to make sense of AI decisions, the strategies they employ when systems behave unexpectedly, and their experiences with control handoffs and feedback mechanisms. Our findings reveal that drivers develop sophisticated folk theories---often using anthropomorphic metaphors describing systems that ``see,'' ``hesitate,'' or become ``overwhelmed''---yet lack informational resources to validate these theories or meaningfully participate in algorithmic governance. We identify contexts where algorithmic opacity manifests acutely, including complex intersections, adverse weather, and rural environments. Current AV designs position drivers as passive data sources rather than epistemic agents, creating accountability gaps that undermine trust and safety. Drawing on critical data studies and algorithmic accountability literature, we propose a framework for participatory algorithmic governance that would provide drivers with transparency into AI decision-making and meaningful channels for contributing to system improvement. This research contributes to understanding how users navigate datafied sociotechnical systems in safety-critical contexts.
\end{abstract}


\section{Introduction}

Autonomous vehicles (AVs) represent a critical site where algorithmic decision-making intersects with high-stakes human activity. The Society of Automotive Engineers (SAE) defines a taxonomy of driving automation ranging from Level 0 (fully manual) to Level 5 (fully autonomous). At Level 2, the vehicle can simultaneously control steering and acceleration, but the human driver must continuously monitor the driving environment and remain ready to intervene at any moment. At Level 3, the vehicle can monitor its surroundings and make situational judgments---such as accelerating past a slow-moving vehicle---but the driver must still be prepared to resume control when the system requests it \citep{reilhac2017user}. As vehicles operating at these intermediate levels become increasingly common, drivers must navigate a complex sociotechnical arrangement in which control shifts dynamically between human and machine \citep{castro2021trends}. Unlike many algorithmic systems that operate invisibly, AV algorithms make observable, consequential decisions in real-time---decisions about speed, lane position, and obstacle avoidance that drivers can directly evaluate against their own expectations and expertise.

This visibility creates a unique context for studying how users make sense of algorithmic systems. When an AV brakes unexpectedly on a highway or hesitates at an intersection, the driver must rapidly construct an explanation for this behavior while deciding whether to intervene. These moments of algorithmic breakdown---when system behavior diverges from user expectations---offer valuable windows into the folk theories people construct to understand opaque computational systems \citep{devito2017algorithms, eslami2016first}.

The opacity of machine learning systems has been documented extensively in critical data studies \citep{burrell2016machine, pasquale2015black}. However, much scholarship has focused on algorithmic systems making decisions \textit{about} people rather than systems making decisions \textit{with} people in shared, real-time tasks. AVs present a distinctive case: drivers are simultaneously subjects of algorithmic surveillance (their behavior is monitored for training), users of algorithmic services (the autopilot assists driving), and potential victims of algorithmic failure (system errors can cause accidents). This tripartite positioning creates complex dynamics around trust, agency, and accountability.

Drawing on 16 semi-structured interviews with AV drivers in the United States, this study investigates: (1) What folk theories do drivers construct to explain algorithmic decision-making in semi-autonomous vehicles? (2) How does algorithmic opacity shape driver strategies, trust, and sense of agency? (3) What mechanisms exist---or are absent---for drivers to participate in algorithmic accountability? We find that while drivers develop rich explanatory frameworks for AI behavior, current AV designs systematically exclude them from meaningful participation in algorithmic governance, positioning them as data sources rather than epistemic agents with valuable contextual knowledge.

\section{Theoretical Framework}

\subsection{Algorithmic Opacity and the Black Box}

The challenge of understanding algorithmic systems has been characterized through the metaphor of the ``black box'' \citep{pasquale2015black}. \citet{burrell2016machine} identifies three forms of algorithmic opacity: intentional corporate secrecy, technical illiteracy among users, and the fundamental inscrutability of machine learning systems whose operations cannot be fully explained even by their designers. All three forms are present in the AV context: manufacturers protect proprietary algorithms; most drivers lack technical expertise; and deep learning systems resist straightforward explanation.

This opacity has significant implications for accountability. As \citet{ananny2018seeing} argue, transparency alone is insufficient---systems must be designed for meaningful accountability enabling affected parties to understand, contest, and shape algorithmic decisions. In safety-critical contexts like driving, the stakes of opacity are particularly high: drivers cannot adequately monitor systems they do not understand, and accountability for accidents becomes diffuse when decision-making processes are inscrutable \citep{elish2025moral}.

The challenge of algorithmic opacity can be understood through \cite{haraway1991simians}'s metaphor of the cyborg, which emphasizes the heterogeneous quality of technology---the invisible welding together of the organic and inorganic, the material and social \cite{book} -- ``we do not know if they are benign or not until we have experienced them. Judgement is suspended until we have constructed knowledge about them''. This observation is particularly salient for AV systems, where drivers must navigate the ambiguity of algorithmic decision-making without clear knowledge of whether system behaviors will prove safe or dangerous until they have been experienced.

\subsection{Datafied Driving and Digital Companion Species}

The integration of AI systems into vehicles represents a broader process of datafication in which driving experiences are transformed through novel data processes. \citet{hind_dashboard_2021} analyzes how the redesign of vehicle dashboards has restructured car-related data, transforming ``vehicle data'' into ``driving data'' through convergence and customization of dashboard features. This datafication occurs through strategic design decisions that persuade drivers to participate in novel practices---such as voice-activated navigation systems that impose new grammars of action, capturing automotive behavior in newly datafied ways. The result is what \citeauthor{hind_dashboard_2021} terms a ``semantic colonization,'' where the very language and concepts drivers use to navigate are reconfigured through algorithmic systems.

This transformation can be understood through \citet{lupton2016digital}'s concept of ``digital companion species,'' drawing on Haraway's work on human-nonhuman assemblages. \citeauthor{lupton2016digital} argues that digital data assemblages are lively entities with which humans co-evolve---learning from each other in ongoing relational encounters. Like companion species, these assemblages are both part of us and beyond our complete control, continuously configured and reconfigured as they circulate through digital data economies. In the AV context, drivers exist in continuous relationship with algorithmic systems that generate, process, and respond to data flows, creating what \citeauthor{lupton2016digital} describes as ``digital data-human assemblages'' characterized by mutual dependency and co-evolution.

\subsection{AI Systems as Distinct from Designed Technology}

Understanding AVs requires recognizing that machine learning-based systems differ fundamentally from traditional designed technologies. \citet{schulz2025generative} argues that while designed technology is developed as predefined means for particular ends, AI systems acquire their features through learning---particularly unsupervised learning. This means that in their pre-trained state, AI models are ``task-agnostic'' and lack the ``toolness'' of designed technology. Consequently, operating these systems cannot follow the same logic as instructing conventional technology; rather, it requires what \citeauthor{schulz2025generative} describes as ``strategic interaction.''

The agency of AI systems, according to \citet{schulz2025generative}, lies in their capability to mobilize machine-learned versions of human experiential knowledge. These systems can be understood as ``agentified human knowledge''---internal representations learned from products of human thought and action that become agentic through the system's capability to generate new content. This fundamentally distinguishes AVs from earlier automotive technologies: drivers interact with systems that reflect learned patterns of human knowledge rather than explicitly programmed sequences of action. This has profound implications for how drivers develop folk theories, as they must make sense of systems whose behaviors emerge from learned patterns rather than designed rules.

This distinction between designed technology and AI systems parallels broader insights about how users appropriate technology in everyday life. As \cite{book} argue, ``consumption is always production'' and users become tinkerers or, in Levi-Strauss's terms, \textit{bricoleurs}---actively constructing meaning and practice around technologies rather than passively accepting their designed purposes. However, AI systems introduce a new dimension to this appropriation: while traditional technologies could be domesticated through understanding their designed functions, AI systems resist such straightforward domestication because their behaviors emerge from learned patterns that may not align with any explicit design intent \citep{schulz2025generative}. This creates what we refer to as a \textit{domestication gap}---a tension between users' need to integrate AI systems into everyday life and the fundamental inscrutability of how those systems generate behavior.

\subsection{Folk Theories of Algorithmic Systems}

When confronted with opaque systems, users construct ``folk theories''---intuitive explanations helping them predict and make sense of system behavior \citep{gelman2011concepts}. Folk theories are informal, intuitive understandings of how algorithmic systems work, grounded in practice and experience rather than technical knowledge \citep{zhao2025boosting}. They are continuously revised as users' perceptions change through ongoing interaction with systems, and serve as behavioral foundations for how users engage with and resist algorithmic arrangements. Research on social media platforms has documented how users develop folk theories about recommendation algorithms, content moderation, and feed curation \citep{devito2017algorithms, eslami2016first, rader2015understanding}. These theories, while often technically inaccurate, serve important functions: they help users feel in control, guide behavioral strategies, and provide frameworks for evaluating system fairness.

\citet{siles2020folk} extend this work by examining folk theories as ``data assemblages''---configurations of users, data, and algorithms enacted differently across contexts. Their study of Spotify users found two primary folk theories: one personifying the algorithm as a social being that surveils users, another viewing it as a trainable machine. These folk theories shaped how users engaged with the platform and perceived their agency. \citet{zhao2025boosting} demonstrate how folk theories not only shape individual behavior but also collective action, documenting how Chinese social media users developed cross-platform folk theories about algorithmic visibility to coordinate collective efforts in boosting content popularity. Their research reveals that within groups engaging in algorithmic resistance, conflicting folk theories can emerge---creating internal power struggles that undermine the consistency and continuity of collective action. This highlights that folk theories are not merely individual cognitive constructs but are negotiated, contested, and collaboratively constructed within communities.

The AV context differs from social media importantly. Folk theories about recommendation algorithms concern relatively low-stakes decisions, while folk theories about AV algorithms concern safety-critical situations. Additionally, AV algorithms operate in real-time physical environments where decisions are immediately observable and consequential, creating both greater urgency for accurate understanding and more opportunities for folk theory formation through direct observation.

Understanding folk theories in the AV context requires attention to how drivers continuously revise their understanding through ongoing interaction. As \cite{book} observe, domestication is not a one-time process but involves continuous negotiation: ``artefacts might then be re-domesticated, even radically'' as practical routines of use are broken, needs change, or persons involved shift. This insight is crucial for AV systems, where drivers must repeatedly update their folk theories as they encounter new edge cases, software updates change system behavior, or their own understanding evolves through accumulated experience. The ``everyday struggles and negotiations'' that \cite{book} identifies as shaping technology are particularly intense in the AV context, where the stakes of misunderstanding can be immediate and severe.

\subsection{Participatory Algorithmic Governance}

Critical scholars have called for moving beyond transparency toward participatory forms of algorithmic governance \citep{katell2020toward}. This involves creating mechanisms through which affected communities can meaningfully shape how algorithmic systems are designed, deployed, and evaluated. \citet{sloane2022participation} identify multiple models of participation, from consultative approaches gathering user feedback to more radical forms of co-design and community ownership.

In the AV context, drivers possess valuable contextual knowledge about local driving conditions, edge cases, and system failures that could improve algorithmic performance. However, current feedback mechanisms are typically limited to passive data collection rather than active user participation. This reflects broader patterns in AI development where user expertise is extracted through data collection but users are excluded from design decisions \citep{gray2019ghost}.

\section{Methods}

\subsection{Research Design}

We conducted 16 semi-structured interviews with drivers who own and regularly use vehicles with advanced driver-assistance systems at SAE Levels 2-3. This sample size allowed for theoretical saturation while enabling in-depth exploration of individual experiences \citep{guest2006how}. Participants were recruited through social media platforms (Reddit, Twitter, Facebook), the Volunteer Science platform, and recruitment flyers at electric vehicle charging stations in the Northeastern United States.

Eligibility criteria required participants to have owned and driven their vehicle for at least six months, ensuring sufficient experience to have developed folk theories and encountered edge cases. Participants completed a screening survey collecting demographic information and details about their vehicle and driver-assistance system.

\subsection{Participants}

Our sample included 13 male and 3 female participants across 10 US states (see Table 1). Educational attainment ranged from high school to doctoral degrees. Twelve participants drove Tesla vehicles (with FSD Beta, FSD, or standard Autopilot), two drove Polestar 2 vehicles with Pilot Assist, one drove a Chevrolet Bolt, and one drove a Mitsubishi Outlander with M-Pilot Assist. This distribution reflects the current market dominance of Tesla in the semi-autonomous vehicle space while capturing variation across manufacturers.

\begin{table}[h]
\centering
\small
\begin{tabular}{llllll}
\toprule
ID & Gender & State & Education & Vehicle & AI System \\
\midrule
P1 & F & WA & Graduate & Chevrolet Bolt & Autopilot \\
P2 & F & WA & Graduate & Tesla Model Y & FSD Beta \\
P3 & M & CA & Bachelor's & Tesla Model 3 & FSD Beta \\
P4 & M & NJ & N/A & Tesla Model 3 & FSD Beta \\
P5 & M & GA & High School & Tesla Model 3 & FSD Beta \\
P6 & M & TX & Bachelor's & Polestar 2 & Pilot Assist \\
P7 & M & NJ & Ph.D. & Polestar 2 & Pilot Assist \\
P8 & M & NJ & Graduate & Tesla Model 3 & FSD Beta \\
P9 & M & MD & Bachelor's & Tesla Model 3 & FSD Beta \\
P10 & M & TN & Graduate & Tesla Model Y & FSD Beta \\
P11 & F & CA & Ph.D. & Tesla Model S/X & FSD \\
P12 & M & VA & Graduate & Tesla Model 3 & FSD Beta \\
P13 & M & WA & High School & Tesla Model Y & FSD Beta \\
P14 & M & TX & High School & Mitsubishi & M-Pilot Assist \\
P15 & M & NJ & Graduate & Tesla Model 3 & FSD \\
P16 & M & NC & Bachelor's & Tesla Model 3 & FSD Beta \\
\bottomrule
\end{tabular}
\caption{Participant demographics}
\end{table}

\subsection{Interview Protocol}

Interviews lasted 45-90 minutes and covered: (1) decision-making around purchasing an AV; (2) initial setup and customization of driver-assistance features; (3) experiences with system behavior that matched or contradicted expectations; (4) strategies for managing unexpected system behavior; (5) experiences with control handoffs; (6) awareness of and engagement with feedback mechanisms.

\subsection{Analysis}

Interviews were transcribed using Otter AI and reviewed for accuracy by the research team, with particular attention to technical terminology and participant descriptions of system behavior. Transcripts were imported into NVivo for analysis. Following \citet{charmaz2006constructing}, we employed constructivist grounded theory, beginning with open coding to identify concepts, then using focused coding to develop analytical categories. This approach is well-suited to studying emergent sense-making practices in sociotechnical contexts \citep{clarke2003situational}, as it allows analytical categories to develop inductively from participants' own accounts rather than imposing predetermined frameworks.

The first author conducted an initial coding pass through all transcripts, attending to detailed descriptions of driving scenarios and developing initial interpretive memos \citep{saldana2011fundamentals}. Both authors then conducted several iterations of focused coding, meeting regularly to discuss emerging patterns and refine analytical categories through constant comparison \citep{charmaz2006constructing}. Following \citet{braun2006using}, we treated themes not as entities passively residing in data but as actively constructed through the researchers' engagement with participants' accounts---an orientation consistent with the reflexive analytical stance common in critical data studies \citep{barad2007meeting, kennedy2016post}.

Our analysis attended specifically to: (1) moments of algorithmic breakdown where system behavior diverged from expectations; (2) the explanatory frameworks participants constructed to interpret these breakdowns; (3) strategies participants developed in response to perceived system limitations; (4) participants' sense of agency and accountability within the human-AI assemblage; and (5) the presence, absence, and adequacy of channels through which drivers could communicate their experiences back to system designers. In attending to these dimensions, we drew on the concept of ``interpretive flexibility'' from science and technology studies \citep{pinch1984social}---recognizing that the same algorithmic behavior could be interpreted differently depending on participants' prior experiences, folk theories, and driving contexts.

\section{Findings}

\subsection{Folk Theories of Algorithmic Perception}

Participants consistently developed folk theories organized around what the AV's AI system could ``see'' or ``sense.'' These theories drew heavily on anthropomorphic metaphors, attributing human-like perceptual capacities and limitations to the algorithmic system.

P15 explained the system's struggles in urban environments through a theory of perceptual overload: ``In cases where there are too many things around and too many variables which are the moving vehicles...people...pedestrians and everything, it senses everything and it will stop and it will break and it will not move.'' Here, the algorithm is imagined as having finite attentional capacity that becomes overwhelmed---a characteristically human limitation projected onto the machine.

Conversely, P7 theorized that the system required sufficient environmental stimulation: ``When you're driving [through] vast stretches in North Dakota and Montana, you can go a long time without seeing a car...there's something about how it calibrates and keeps itself going. And if it doesn't have enough input, it just can't keep going.'' This folk theory frames the algorithm as requiring environmental nourishment.

P13 attributed sharp limits to the system's perception: ``It would try to stop in the middle of the road at the crest of a hill because it couldn't see that there was road over there.'' The metaphor of ``seeing'' allowed P13 to predict system behavior while explaining why human oversight remained necessary.

These perceptual folk theories served pragmatic functions. By constructing models of what the system could perceive, participants developed strategies for when to trust the autopilot. P5 reported avoiding autopilot in ``heavy tourist situations where you're driving down, say a boardwalk...think like a wedding every 500 feet sort of day.'' His theory of the system's perceptual limitations directly shaped his driving behavior.

Participants also theorized about the granularity of the system's perception---what kinds of objects it could and could not detect. P10 developed a size-based theory: ``A deer ran out in front of me, the vehicle spotted it, it swerved out of the way. But I've also had large...boxes in the road or large potholes in the road, and the vehicle would just drive through.'' P9 refined this further, observing that the car would not ``avoid anything that does not look like a cone.'' These theories suggest that participants imagined the algorithm as having a categorical perception system---a library of recognized objects against which environmental stimuli were matched. Objects falling outside this imagined library were understood to be invisible to the system. P12 offered a complementary view, suggesting the system could ``extend their senses'' beyond what was available to a human driver alone, for instance detecting hidden pedestrians when pulling out of a driveway. This folk theory positioned the algorithm as possessing perceptual \textit{capabilities} exceeding human ability in some respects while remaining inferior in others---a nuanced understanding of algorithmic capacity that shaped context-specific trust decisions.

\subsection{Folk Theories of Algorithmic Cognition}

Beyond perception, participants developed theories about how the system ``thought'' or ``decided.'' These often emerged from observations of hesitant or erratic behavior at decision points.

P4 described the system's behavior at complex intersections: ``It would be trying to figure out which way to go. And sometimes it would commit, it would look like it was committing to the wrong road.'' The language of ``figuring out'' and ``committing'' attributes deliberative cognitive processes to the algorithm. P3 similarly noted that the AV ``would get jerky on traffic circles,'' suggesting the system experienced confusion or indecision.

Several participants theorized about the algorithm's ``personality.'' Tesla's autopilot offers three driving profiles (Chill, Average, and Assertive/``Mad Max''), and participants constructed theories about what distinguished them. P4 found that ``Aggressive'' mode ``actually does seem to be driving more like a human because, you know, people...roll through stop signs all the time.'' P11, by contrast, found ``Chill'' mode ``way too conservative...it will not just slow down for the speed-bump, it will come to a complete stop.''

These theories reveal an important tension: participants wanted the algorithm to drive ``like a human''---meaning in accordance with local driving norms---but the system appeared to operate according to legal rules or opaque optimization criteria not matching human driving patterns. This mismatch between algorithmic behavior and norms created situations where technically ``safe'' decisions (full stops) could create actual danger (rear-end collisions from following drivers).

\subsection{Strategies for Managing Algorithmic Opacity}

Participants developed various strategies for navigating algorithmic uncertainty, ranging from calibration practices to defensive driving techniques assuming system failure.

\textbf{Gradual calibration of aggressiveness settings.} P16 described gradual calibration: ``My first drive was long enough for it to calibrate the autopilot...I was extremely cautious because I didn't trust it...after I became more familiarized with the software...over time [I] started [increasing] the aggressiveness.'' {P10 explicitly customized settings to match perceived driving conditions: ``I changed the lane change to \textit{Mad Max}. And I changed the highway version to \textit{Assertive}, I changed some of my speed profiles, instead of just set the speed limit. I changed it to 5\% over.'' This account frames the driver-AI relationship as requiring mutual adjustment ---drivers developing folk theories about which settings align with their driving contexts and risk tolerances. However, these adjustments occur without transparency about what the different modes actually optimize for, forcing drivers to learn through trial and error what ``Mad Max'' or ``Assertive'' mean in practice.

\textbf{Navigating safety-risk trade-offs in settings.} P11's account revealed how settings adjustments involved navigating complex safety trade-offs shaped by local driving norms. She explained that while Tesla blamed an accident on a driver setting the following distance to ``one'' (the most aggressive setting), local driving conditions made this setting practically necessary: ``You have to set following to be one in this area because otherwise people are cutting in front of you all the time...everyone sets their up follow to the one.'' This illustrates a tension between manufacturer safety guidelines and situated driving knowledge---the ``safest'' setting according to the algorithm created hazards in contexts where other drivers interpreted large following distances as opportunities to merge. P11 also disabled lane departure warnings that vibrated or beeped when the vehicle approached lane markings: ``The lanes are often kind of spaced, assuming people are going to cut the corner a little bit....So, I disabled that because that was annoying as heck.'' What the system coded as deviation requiring correction, the driver understood as normal driving practice adapted to local road geometry.

\textbf{Defensive vigilance strategies.} P13 described a more defensive stance requiring constant vigilance: ``You're like a 100th of a second away from grabbing it all the time. It was exhausting to drive. Because you had to be there.'' For P13, the opacity of the system meant trust was impossible; the only safe strategy was assuming failure could occur at any moment. This strategy reflects what \citet{parasuraman2008humans} identify as the ``automation paradox'': systems requiring continuous monitoring for rare failures can induce fatigue and reduced situational awareness, potentially increasing rather than decreasing risk.

\textbf{Context-specific rules through trial and error.} Several participants developed context-specific rules. P12 learned that leaving a highway for rural roads required intervention: ``That puppy doesn't slow down as fast as it should...it should read the 35 mile an hour speed limit coming up.'' P4 developed rules about when the system's conservative programming became counterproductive: ``The vehicle would...flow more smoothly...if they relaxed some of those precautions'' for situations like rolling through stop signs ``as long as there's no other cars around.'' Through trial and error, participants mapped the system's limitations onto geographic and situational categories, developing what might be termed ``negative folk theories''---knowledge about where the system fails rather than how it succeeds.

\textbf{Vicarious learning through media and community.} Notably, some strategies were shaped not only by personal experience but by information circulating through media and community networks. P11 reduced her use of certain autopilot features not because of direct experience but because of a news report about an accident in which the manufacturer attributed fault to the driver's aggressive settings. This prompted P11 to reconsider her own configuration choices---a form of vicarious folk theorizing mediated through public discourse rather than personal algorithmic encounter. Her account illustrates how folk theories of AV algorithms are constructed not only through individual interaction but through what \citet{couldry2019costs} describe as broader ``data relations''---the social contexts through which people come to understand their positioning within datafied systems.

\textbf{Embodied discomfort prompting intervention.} P12's calibration strategy was shaped by social considerations: he initially selected ``Chill'' mode ``thinking that would...be more comfortable for my wife,'' but found that the mode's smooth speed maintenance on ramps ``was making him seasick.'' The angular momentum on curves became physically uncomfortable, leading P12 to take manual control during transitions. P11 similarly found that ``Chill'' mode was ``way too conservative...it will not just slow down for the speed-bump, it will come to a complete stop''---behavior that, while theoretically safer, could paradoxically cause accidents when following drivers expected normal deceleration patterns. These examples illustrate how algorithmic opacity extends beyond cognitive uncertainty to embodied, visceral experience---the algorithm's behavior produced physical discomfort that could not be resolved through better understanding but only through manual override.

\subsection{Algorithmic Opacity in Specific Contexts}

Participants identified recurring contexts where algorithmic opacity manifested acutely:

\textbf{Complex intersections and roundabouts.} P3 reported the AV ``would get jerky on traffic circles,'' struggling to follow the painted arrows on roundabouts. P4 noted the vehicle would be ``trying to figure out which way to go'' at complex junctions connecting more than two roads. He described how the vehicle would ``commit, it would look like it was committing to the wrong road,'' creating situations where he would ``usually take over and...get through the intersection like a human would.'' These multi-directional decision points revealed the limitations of systems trained primarily on simpler road geometries.

\textbf{Unprotected left turns.} Multiple participants identified unprotected left turns---where oncoming vehicles have right of way---as particularly challenging for the autopilot. P4 explained: ``There are some unprotected left turns that [the vehicle]...is hesitant on and I think taking out some of those, whatever algorithms they put in to be to be extra cautious...if they relaxed some of those precautions...the vehicle would...flow more smoothly and it wouldn't be so jerky and hesitant.'' P4 further noted the autopilot ``does not know [how to] respond to aggressive drivers'' when attempting left turns or in intersections. This revealed a fundamental tension: the system's conservative programming, designed for safety, produced driving behavior that deviated from human norms in ways that could paradoxically create hazards.

\textbf{Construction zones.} P11 described how the car ``glitch[es]...maybe the lane markings are not very [clear]...there's some construction [where] the lane markings went away.'' P3 provided a vivid account of a construction site where ``the line was literally on the left side up against the...concrete barrier. And I mean, the line actually went under the barrier every once in a while, because it was a construction site...it's like, holy shit and you grab the steering wheel and move it away from the barrier... if you see the lines going underneath the concrete barrier, be assured that you need to take control!'' These construction scenarios exposed the system's dependence on consistent lane markings, transforming routine highway maintenance into opacity-inducing environments where the algorithm's perceptual framework broke down. P11 described how in such situations, the car would often ``just be like, nope, your turn,'' abruptly returning control to the driver precisely when the driving environment had become most challenging.

\textbf{Adverse weather.} P2 noted poor performance in fog, attributing this to limitations in the system's light-based sensing. P10 observed: ``I think it was more just based on the density of the rain...a heavy rainstorm is what caused it.'' P14 developed a theory linking weather to the system's reliance on visual lane markers: the autopilot ``is disengaging when it loses where it's centering in the road, right? It's using the line so it'll lose its thing.'' P11 noted that extreme heat also affected performance: ``If it's really hot out, sometimes the centers [(yellow lines)] are a little wonky.'' Multiple participants described ``phantom braking''---sudden deceleration for no apparent reason---which proved particularly distressing because it resisted folk theorizing. P13 captured this interpretive frustration: ``I don't really understand what triggers it because we look around and it's like now send it in. But I don't know what is making it do that. It's some kind of maybe it's getting an optical illusion from the front camera that it's going down into a hole or something. But yeah, we haven't really figured out the latest bout.'' The phrase ``the latest bout'' is revealing: it suggests phantom braking arrives in episodic waves---perhaps correlated with software updates---which further complicates folk theorizing by introducing temporal variability into an already opaque system. P16 connected phantom braking to safety concerns in winter conditions, noting he does ``not trust autopilot as well in the snow as a human due to the fact of the phantom braking. If it is actually a slick road, it's probably going to cause someone to lose traction, because it would break harder than a human would.''

\textbf{Speed management across contexts.} Several participants identified a distinct category of opacity around the algorithm's speed regulation. P11 described unexpected braking at highway speeds: ``I'm going 75 on the freeway, and [the vehicle is] suddenly breaking. And there's traffic behind me, right? Like that is a very unsafe thing to do.'' She suspected the cause was shadows cast by overpasses---a folk theory attributing the system's behavior to an optical misperception. P14 described the inverse problem: the autopilot would gradually decelerate without apparent cause, driving at 55 miles per hour on a 75 mile-per-hour highway. These accounts reveal that opacity is not limited to dramatic failures but also pervades mundane speed adjustments where the algorithm's criteria for acceleration and deceleration remain invisible to the driver.

\textbf{Rural environments.} P13 described ``all these bridge transitions and spots in Montana where the snowplows have just ripped out...sections of road. The car goes right through that.'' P16's autopilot would jerk at the gravel road leading to his home ---a recurring issue that generated valuable contextual knowledge about edge cases yet found no structured channel for system improvement.

\textbf{Context-dependent usability.} Beyond specific physical environments, participants identified patterns in when they chose to engage autopilot. Multiple participants reported comfort using autopilot for monotonous, repetitive tasks---particularly daily highway commutes where conditions were familiar. P13 described highway driving where ``yeah, now you can sing and dance and eat dinner in there. As long as you're hanging on to the steering wheel and looking down the road.'' However, trust evaporated in ``heavy tourist situations where you're driving down, say a boardwalk...think like a wedding every 500 feet sort of day and people just crossing the street'' (P5). This suggests drivers develop nuanced understandings of contexts where the system performs reliably versus situations requiring human judgment---yet these situated assessments remain individual rather than collective, preventing the aggregation of contextual knowledge that could improve system design.

These contexts share a common feature: they deviate from the ``standard'' driving environments on which the AI was presumably trained, revealing the limits of the system's training data.

\subsection{Handoffs and the Distribution of Agency}

Control handoffs---moments when the system returns control to the driver---proved particularly revealing of accountability structures. Participants described three distinct experiences.

Some reported handoffs as predictable. P2 stated: ``I never experienced [the car handing over control] for no reason. If it's ever disengaged, it's because...it doesn't have the appropriate input.'' P2 had constructed a folk theory making handoffs interpretable.

Others experienced handoffs as anxiety-producing. P10 described being ``slightly scary'' while acknowledging risk acceptance: ``I understand...using full self-driving in a beta format, that it is still a beta.'' P11 articulated broader concerns: ``I read enough about the autopilot disengaging basically right before an accident for a lot of people...it just like, dumps me out of autopilot whenever things get slightly confusing.'' P13 provided a particularly vivid account of handoff anxiety on rural roads: ``A back-country road and you hit that 90 degree turn from a fairly good speed, it would kick out if it couldn't see around the corner. And they would hand it back to you...if you're not paying attention...yeah, you could theoretically drive off a cliff.'' This scenario reveals a fundamental paradox in the distribution of agency: the system relinquishes control precisely at moments when the driver's situational awareness is most likely to be diminished---having relied on the autopilot, the driver may not have been tracking the approaching turn with the vigilance required for manual driving. P10 recounted one unexpected handoff on a ``six lane [highway with] some turnouts...and heavy traffic...[the] damn thing turned off.''

A third category involved inadvertent handoffs triggered by the driver. P13 noted: ``Most of the time [the transition] is accidental...if I head into a corner...the reflexes to tap the brake and that shuts it off.'' He elaborated on the physical challenge of maintaining the required steering wheel pressure without triggering a takeover: ``I leave my elbow on the armrest, right, so that I can keep a constant tension on the steering wheel because you got to have certain amount of pressure on the steering wheel so it knows you're there...if you flinch, you grab the wheel and if you grab the wheel...[you take control]...it's yours. Oh, I didn't want it.'' P16 concurred: ``It takes as little as just barely tapping...your foot touches the brake pedal and it disengages.'' Decades of driving experience had created embodied responses conflicting with the interaction paradigm required by semi-autonomous systems. These inadvertent handoffs illustrate a mismatch between the algorithm's binary model of control (engaged/disengaged) and the continuous, embodied nature of human driving practice.

These experiences reveal fundamental ambiguities in agency and responsibility distribution. When the system ``dumps'' control at moments of ``confusion,'' accountability becomes murky: was the human adequately monitoring? Did the system provide sufficient warning?

\subsection{Feedback Mechanisms and Exclusion from Algorithmic Governance}

Our interviews revealed that meaningful channels for driver participation in algorithmic improvement were largely absent. This absence was itself a significant finding: across 16 interviews with experienced AV drivers, most participants had never attempted to provide structured feedback about system failures, and those who had encountered significant opacity about whether and how their input was used.

\subsubsection{The snapshot as opaque feedback channel.} P16 was the only participant who described actively using a built-in feedback mechanism: ``[I] hit a button [which] takes a little snapshot recording. And when it takes that snapshot recording, it sends that data to the engineers.'' However, P16's account reveals multiple layers of opacity surrounding this mechanism. He was ``not sure what happened once the snapshot was taken. There were no signals to the driver explaining what data was shared.'' When the system's handling of a problematic gravel road later improved, P16 attributed this to his repeated feedback but could not confirm a causal link: ``I guess just over time...every drive me just hitting that eventually, an update came out.'' The uncertainty in P16's account---``I guess,'' ``eventually''---reveals how even the most engaged driver-participant was left to construct folk theories not only about system behavior but about the feedback process itself. The feedback mechanism, much like the driving algorithm, operated as a black box \citep{pasquale2015black}.

\subsubsection{Informal feedback and workaround strategies.} In the absence of formal feedback channels, several participants developed informal strategies that functioned as surrogate forms of communication with the system. P2 expressed frustration about the lane-change feature's opacity, noting that ``it will be nice if they [the manufacturers] could provide a better depiction of how far it's crawling.'' This statement encodes both a desire for transparency and an implicit wish for a channel through which such design preferences could reach developers. P4 articulated a similar desire, suggesting that ``if they relaxed some of those precautions...the vehicle would...flow more smoothly and it wouldn't be so jerky and hesitant.'' Both participants had clear, specific suggestions for algorithmic improvement but no avenue through which to communicate them.

P13's reaction to unexplained system behavior captured the interpretive void created by absent feedback infrastructure: ``Every road trip we take, this car does something to make you go. Did that just happen?!'' The rhetorical question directed at no one in particular---neither the car nor the manufacturer---illustrates what we term \textit{communicative isolation}: drivers accumulated rich experiential knowledge about system failures but had no structured recipient for this knowledge. Their observations remained private frustrations rather than becoming inputs to algorithmic improvement.

\subsubsection{Data access and informational asymmetry.} The asymmetry between what manufacturers extracted from drivers and what drivers could access about system behavior was starkly illustrated by the issue of vehicle data logs. We asked all participants whether they had access to their vehicle's data logs. P11 was the only respondent who shared Tesla logs with us, and only because she resided in California, where state privacy regulations granted residents the right to request their data. Other Tesla-owning participants were unable to access their logs. P11 herself had not accessed these logs before our inquiry, suggesting that even where access rights existed, the practical infrastructure for drivers to engage with their own data was effectively absent.

This data access barrier illustrates what \citet{couldry2019costs} describe as ``data relations''---the structured asymmetries through which some actors gain knowledge from data while others are reduced to data sources. Drivers continuously generate telemetry data through their vehicles' sensors---data that feeds machine learning systems and informs algorithmic updates. Yet this extraction proceeds unidirectionally: driver behavior flows to manufacturers as training data, but information about algorithmic functioning does not flow back in usable forms. P13's bewilderment about phantom braking captures this asymmetry precisely: ``I don't really understand what triggers it because we look around and it's like...I don't know what is making it do that.'' The system harvests P13's driving data while providing no legible account of its own decision-making.

\subsubsection{News media and community forums as surrogate accountability.} Absent direct channels, some participants turned to external information sources to make sense of algorithmic behavior. P11's decision to modify her settings was prompted not by manufacturer communication but by a news report about an accident attributed to aggressive settings. The news media thus functioned as an indirect, partial, and unreliable accountability mechanism---one that provided cautionary narratives without the technical specificity needed to inform driving decisions. However, P11's response to this news report reveals the inadequacy of media-based accountability. After learning about the accident, she faced a dilemma: the manufacturer blamed the crash on the driver setting following distance to one, positioning higher settings as safer. Yet P11 explained that in her driving environment, 'you have to set following to be one in this area because otherwise people are cutting in front of you all the time. [They] think there's a big gap in front of you, like everyone sets their [following distance] to one.' The news report thus created confusion rather than clarity—framing a setting as recklessly aggressive while local driving conditions made it practically necessary. This exemplifies how surrogate accountability through news media operates without crucial contextual knowledge: drivers receive cautionary narratives about system misuse but lack manufacturer guidance about how to balance competing safety considerations in specific environments. The media reports system failures, but cannot provide the technical or contextual specificity needed for drivers to calibrate their own practices.

P7 drew on a general understanding of machine learning to theorize about system limitations: ``I guess it's a machine learning device. So, it needs constant input to be able to adjust.'' This folk theory, while technically imprecise, represents an attempt to mobilize publicly available knowledge about AI systems to compensate for the manufacturer's failure to provide explanation. P7's folk theory emerged from experiences driving through rural Montana and North Dakota, where 'you can go a long time without seeing a car. And so, there's something about there must be something about how it calibrates and keeps itself going. And if it doesn't have enough input, it just can't keep going.' This longer quote reveals how P7 constructed his machine learning theory from embodied experience of rural driving—noticing the system struggled in low-stimulus environments and theorizing that 'constant input' was necessary for algorithmic functioning. His theory, while technically imprecise about how machine learning systems actually operate, demonstrates sophisticated reasoning from observation to explanation, mobilizing publicly available AI discourse to compensate for manufacturer opacity.

These surrogate channels---news reports, online communities, general AI literacy---highlight the informational labor that drivers must perform in the absence of structured transparency mechanisms. As \citet{kennedy2016post} argue in their analysis of data literacy, the burden of interpretation falls disproportionately on those with the least access to relevant information. In the AV context, drivers who are most affected by algorithmic decisions are systematically excluded from the information needed to evaluate those decisions. The reliance on these surrogate channels---partial news narratives and general AI literacy---illustrates the \textit{domestication gap} introduced earlier: drivers attempt to domesticate opaque AI systems through bricolage, piecing together fragments of public knowledge, yet these external sources cannot bridge the fundamental inscrutability of how algorithmic decisions emerge from learned patterns rather than designed rules.

\subsubsection{The absence of collective visibility.} A final dimension of feedback exclusion concerns the isolation of individual driver experiences. Each participant described system failures as singular, personal events. No participant was aware of whether other drivers encountered similar issues in similar contexts. P11's overpass-shadow braking, P13's phantom braking ``bouts,'' P16's gravel-road difficulties---each was experienced as an individual confrontation with an opaque system. The absence of mechanisms for collective visibility meant that systemic patterns remained invisible to the driver community even as the manufacturer, with access to fleet-wide data, could presumably identify them. This asymmetry in pattern recognition constitutes a significant dimension of what \citet{ananny2018seeing} term the ``limits of seeing''---not merely the inability to see into an algorithmic system, but the structural prevention of seeing \textit{across} individual experiences to identify systemic failures.

\section{Discussion}

\subsection{From Opacity to Interpretive Labor}

Our findings extend critical data studies scholarship on algorithmic opacity by documenting the interpretive labor users perform when navigating opaque systems in real-time, safety-critical contexts. Unlike recommendation algorithms operating in the background, AV algorithms demand immediate interpretation and response. This creates what we term ``situated folk theorizing''---the rapid construction of explanatory frameworks in response to observable algorithmic behavior.

The folk theories our participants constructed served multiple functions beyond prediction. They provided emotional reassurance, guided behavioral strategies, and enabled accountability narratives. However, the technical inaccuracy of many folk theories raises concerns. When P15 attributes urban struggles to ``sensing everything,'' this may reflect anthropomorphic projection rather than actual system architecture. Such misunderstandings could lead to misplaced trust or inappropriate intervention strategies.

The nature of these folk theories reflects the distinctive character of AI systems as fundamentally different from traditional designed technologies. As \citet{schulz2025generative} argues, machine learning systems acquire their features through learning rather than explicit programming, making them ``task-agnostic'' in their pre-trained state and resistant to the instructional logic that governs conventional tools. This helps explain why our participants' folk theories so frequently deployed anthropomorphic metaphors---systems that ``see,'' ``hesitate,'' ``get overwhelmed.'' Drivers are attempting to make sense of what \citeauthor{schulz2025generative} terms ``agentified human knowledge'': internal representations learned from patterns of human action that generate new behaviors. The folk theory P13 developed about phantom braking---``maybe it's getting an optical illusion''---reflects an intuitive grasp that the system operates through learned pattern recognition rather than rule-following, even as the specific mechanism remains opaque. Understanding AVs thus requires what \citeauthor{schulz2025generative} calls ``strategic interaction'' rather than instrumental use---a shift that current vehicle designs do little to support or scaffold.

\subsection{Data Assemblages and Asymmetric Participation}

Drawing on \citet{siles2020folk}, we can understand the AV as a data assemblage in which drivers, vehicles, algorithms, manufacturers, and regulators are configured in particular relationships. The very language of "autonomous" vehicles obscures these configurations. Despite popular terminology suggesting self-sufficiency, these vehicles are fundamentally networked entities embedded in vast sociotechnical systems \citep{winner1978autonomous}. The rhetoric of autonomy masks the extensive human labor, infrastructural dependencies, and collective data flows that enable operation \citep{bradshaw2013seven}. A Tesla, purchased as an individual consumer object, immediately enrolls its owner into a distributed learning network where driving data flows continuously to the manufacturer. The vehicle's capabilities emerge not from isolated on-board intelligence but from the aggregated experiences of thousands of sibling machines—each driver unwittingly serving as a trainer for algorithmic systems they cannot inspect or influence \citep{stilgoe2018machine}.

This relationship exemplifies what \citet{lupton2016digital} describes as ``digital data-human assemblages,'' configurations in which humans and algorithmic systems exist in continuous, mutually constitutive relationship. \citeauthor{lupton2016digital}'s concept of ``digital companion species'' is particularly apt for AVs: like companion species, these systems are simultaneously ``part of us and beyond our complete control,'' configured and reconfigured through ongoing interaction. The folk theories our participants developed reflect this co-evolutionary dynamic---drivers learning to interpret algorithmic behavior even as their actions generate training data that shapes future system responses. However, as \citeauthor{lupton2016digital} notes, these assemblages circulate through ``digital data economies'' that structure relationships asymmetrically. In the AV context, this asymmetry manifests in manufacturers' capacity to aggregate and analyze fleet-wide patterns while individual drivers remain isolated, unable to validate their folk theories against collective experience.

The absence of collective visibility mechanisms contrasts sharply with user communities documented by \citet{zhao2025boosting}, where social media users developed shared folk theories to coordinate algorithmic resistance. \citeauthor{zhao2025boosting}'s analysis of Chinese content creators reveals how folk theories become ``behavioral foundations'' for collective action, with users developing cross-platform strategies to boost visibility. While \citeauthor{zhao2025boosting} also document conflicts within communities---where competing folk theories undermine collective action---the mere existence of spaces for negotiating shared understanding represents a form of participation absent from AV contexts. Current AV designs prevent the formation of driver communities that could validate, refine, or contest folk theories collectively. P11's adjustment of following distance settings based on news of another driver's accident illustrates the potential for collective knowledge-building, yet such knowledge currently circulates through informal channels rather than structured mechanisms for shared learning.

Our analysis reveals that current configurations systematically position drivers as data sources rather than data partners. Drivers contribute continuous streams of behavioral and environmental data through vehicle sensors. The pressure to ``feed the machine'' reflects what \cite{bates2025feeding} describe as  as AI's ``data dilemma.'' While these data train machine learning systems powering autopilot features, drivers are excluded from decisions about data collection, model training, and system updates. The relationship is extractive: driver knowledge and experience flow to manufacturers, but information about system functioning does not flow back in usable forms.

Yet despite being positioned as passive data sources, drivers are engaging in precisely the kind of active appropriation that \cite{book} identify in their analysis of technology domestication. As Lie and Sorensen argue, "consumption is always production"—users are not passive recipients of technology but bricoleurs who actively construct meaning and practice around technological artifacts. The folk theories documented in our study exemplify this bricolage: drivers piece together explanations from behavioral observations, news reports, online communities, and accumulated experience, creating workable mental models despite manufacturers' opacity. P13's theory about optical illusions causing phantom braking, P15's notion of the system becoming "overwhelmed" in urban environments, P16's understanding of gravel roads as outside the training data—each represents an act of productive consumption, transforming an opaque system into something interpretable through creative synthesis of available resources. However, the \textit{domestication gap} we identified earlier remains: while drivers engage in this interpretive labor, current AV designs neither recognize nor harness it. The productive capacity Lie and Sorensen describe—users' ability to generate valuable knowledge through everyday engagement with technology—is systematically excluded from feeding back into system improvement, rendering drivers' bricolage invisible to the very systems they are working to understand.

This asymmetry has implications for algorithmic accountability. Drivers possess contextual knowledge that could improve system performance—knowledge about local conditions, regional driving norms, and edge cases. P16's experience with the gravel road illustrates this: his repeated encounters generated exactly the kind of situated, contextual feedback that could improve the system. Yet current designs capture this feedback only incidentally rather than creating structured channels for driver participation. The purported "autonomy" of these vehicles thus operates in tension with their fundamental dependence on collective human experience—a dependence that benefits manufacturers while rendering individual drivers' contributions invisible and uncompensated.

\subsection{Toward Participatory Algorithmic Governance}

We propose that AV manufacturers move toward participatory algorithmic governance—designing systems that position drivers as partners in ongoing improvement rather than passive subjects of algorithmic decision-making. This reorientation responds to what \citet{crawford2016can} identify as the fundamental power asymmetries embedded in big data systems, where those who generate data rarely benefit from or control its use. Our proposals also build on \citet{kennedy2016post} call for "data justice" approaches that center the interests of affected communities rather than treating them as mere inputs to optimization processes. This would involve several components.

\textbf{Transparency mechanisms} helping drivers understand why the system made specific decisions. When the AV brakes unexpectedly, drivers could access information about what sensor inputs triggered the response, moving beyond folk theories constructed from behavioral observation alone. This aligns with \citet{ananny2018seeing} argument that meaningful transparency requires not merely opening black boxes but creating conditions for genuine understanding and contestation. As they note, transparency efforts often fail when they provide access to information without the interpretive resources needed to make that information actionable.

\textbf{Structured feedback channels} enabling drivers to contribute contextual knowledge about system failures. Rather than opaque "snapshot" buttons, systems could prompt drivers to categorize incidents, describe contributing factors, and track whether similar situations improve. This approach draws on \citet{gray2019ghost} documentation of how platform companies extract value from user knowledge while obscuring the human labor that sustains algorithmic systems. Structured feedback would make driver contributions visible and valued rather than rendering them, in \citeauthor{gray2019ghost}'s terms, "ghost work" that disappears into corporate data infrastructures.

\textbf{Collective visibility} allowing drivers to see patterns across the user community. If a particular intersection consistently triggers phantom braking across multiple vehicles, this information could be surfaced, enabling collective pressure for improvement and validating individual experiences. Such collective visibility responds to concerns raised by \citet{eubanks2018automating} about how algorithmic systems individualize what are often systemic problems, preventing affected users from recognizing shared experiences and organizing collectively. \citet{beraldo2019data} similarly argue that data activism requires making visible the patterns that algorithmic systems produce across populations.

Drawing on \citet{plantin2018re}'s analysis of data sharing platforms in scholarly infrastructure, we can envision collective visibility mechanisms that leverage platform logic to connect distributed experiences of AV users. \citeauthor{plantin2018re} demonstrate how platforms like Figshare use API architectures to re-integrate scattered components of research infrastructure, enabling data sharing and collective knowledge building across institutions. Similarly, AV manufacturers could develop platform-based systems allowing drivers to aggregate and analyze incidents across the user community---not merely as passive data sources, but as active contributors to collective knowledge about system performance. However, \citeauthor{plantin2018re} also caution that such platforms risk becoming new centralizing entities, potentially splintering knowledge infrastructures in ways that affect equity of service. Any collective visibility mechanism must therefore balance the benefits of aggregation against risks of re-centralization, ensuring that knowledge flows benefit affected communities rather than extracting value for corporate optimization.

\textbf{Accountability mechanisms} clarifying responsibility in human-AI handoffs. Current designs create ambiguity about responsibility when failures occur. Clearer protocols with adequate warning time and explicit confirmation could better distribute agency and accountability. This addresses what \citet{elish2025moral} terms the "moral crumple zone"—the tendency for human operators to absorb blame for system failures even when those failures stem from design decisions beyond their control. In the AV context, drivers currently occupy this crumple zone, expected to monitor systems they cannot understand and intervene in situations they cannot anticipate.

The challenge of establishing accountability is complicated by the platform-based infrastructure through which many AV systems operate. \citet{plantin2018re} examine how data sharing platforms in research contexts raise fundamental questions about institutional responsibility, corporate longevity, and openness. When critical functions---such as safety incident reporting and system performance tracking---are delegated to platform intermediaries, accountability becomes distributed across corporate entities whose goals may not align with driver safety or public interest. \citeauthor{plantin2018re} warn that platforms displacing traditional infrastructures risk creating dependencies on private firms ``whose longevity, openness, and corporate goals remain uncertain.'' For AV systems, this suggests the need for accountability frameworks that extend beyond individual manufacturers to encompass the entire data infrastructure---including cloud services, mapping platforms, and connectivity providers---that enables contemporary autonomous driving. Such frameworks must address not only technical handoff protocols but also the governance structures determining how driver feedback is collected, analyzed, and acted upon within complex corporate ecosystems.

These proposals align with broader calls for algorithmic accountability \citep{diakopoulos2015algorithmic} while attending to the specific constraints of real-time, safety-critical systems. They also respond to \citet{couldry2019data} critique of "data colonialism," in which user-generated data flows extractively toward corporate centers without reciprocal benefit. The goal is not full transparency of proprietary algorithms—which, as \citet{burrell2016machine} notes, may be impossible given the fundamental inscrutability of some machine learning systems—but meaningful accountability enabling affected parties to understand, evaluate, and shape system behavior. Following \citet{katell2020toward}, we envision participatory algorithmic governance not as a one-time consultation but as an ongoing relationship in which drivers' situated knowledge continuously informs system development.

\section{Limitations and Future Directions}

Our study has limitations. Interview methods may introduce recall bias. Future research could employ ride-along observations or diary studies to capture folk theorizing as it occurs. Our sample, while geographically distributed, was predominantly male and Tesla-focused, reflecting demographics of early AV adopters. As semi-autonomous features become standard across manufacturers, research should examine how folk theories vary across populations.

The rapid evolution of AV technology means our findings represent a particular moment. Some issues participants described may be addressed in future updates. However, the fundamental dynamics we document---algorithmic opacity producing folk theorizing, asymmetric participation in data assemblages, accountability gaps in handoffs---are likely to persist as long as AV systems rely on opaque machine learning and position drivers as data sources rather than governance partners.

\section{Conclusion}

This study examined how drivers of semi-autonomous vehicles construct folk theories to navigate algorithmic opacity in safety-critical contexts. We found that while drivers develop sophisticated explanatory frameworks, current AV designs systematically exclude them from meaningful participation in algorithmic governance.

Our findings contribute to critical data studies by documenting the interpretive labor required to navigate opaque AI systems in real-time physical environments. They also have practical implications for AV design: creating mechanisms for transparency, feedback, and collective visibility could improve both system performance and user trust.

The drivers in our study demonstrated remarkable capability for constructing workable folk theories despite significant informational disadvantage. Rather than viewing this as a problem to be solved through better explanation, we suggest it points toward an opportunity: drivers are already engaged in sense-making work that could, with appropriate design, contribute to system improvement. The question is whether manufacturers will design systems that harness this distributed expertise or continue treating drivers as passive subjects of algorithmic governance.




\bibliography{referencesbds}

@article{ananny2018seeing,
  author = {Ananny, Mike and Crawford, Kate},
  title = {Seeing without knowing: Limitations of the transparency ideal and its application to algorithmic accountability},
  journal = {New Media \& Society},
  volume = {20},
  number = {3},
  pages = {973--989},
  year = {2018}
}

@article{burrell2016machine,
  author = {Burrell, Jenna},
  title = {How the machine `thinks': Understanding opacity in machine learning algorithms},
  journal = {Big Data \& Society},
  volume = {3},
  number = {1},
  pages = {1--12},
  year = {2016}
}

@article{castro2021trends,
  author = {Castro, Afonso and Silva, Filipe and Santos, Vitor},
  title = {Trends of human-robot collaboration in industry contexts: Handover, learning, and metrics},
  journal = {Sensors},
  volume = {21},
  number = {12},
  pages = {4113},
  year = {2021}
}

@book{charmaz2006constructing,
  author = {Charmaz, Kathy},
  title = {Constructing Grounded Theory: A Practical Guide Through Qualitative Analysis},
  publisher = {SAGE},
  address = {London},
  year = {2006}
}

@inproceedings{devito2017algorithms,
  author = {DeVito, Michael A. and Gergle, Darren and Birnholtz, Jeremy},
  title = {``Algorithms ruin everything'': \#RIPTwitter, Folk Theories, and Resistance to Algorithmic Change in Social Media},
  booktitle = {Proceedings of the 2017 CHI Conference on Human Factors in Computing Systems},
  pages = {3163--3174},
  year = {2017},
  address = {Denver, CO, USA},
  publisher = {ACM},
  location = {New York}
}

@article{diakopoulos2015algorithmic,
  author = {Diakopoulos, Nicholas},
  title = {Algorithmic accountability: Journalistic investigation of computational power structures},
  journal = {Digital Journalism},
  volume = {3},
  number = {3},
  pages = {398--415},
  year = {2015}
}

@inproceedings{eslami2016first,
  author = {Eslami, Motahhare and Rickman, Aimee and Vaccaro, Kristen and Aleyasen, Amirhossein and Vuong, Andy and Karahalios, Karrie and Hamilton, Kevin and Sandvig, Christian},
  title = {``I always assumed that I wasn't really that close to [her]'': Reasoning about Invisible Algorithms in News Feeds},
  booktitle = {Proceedings of the 2016 CHI Conference on Human Factors in Computing Systems},
  pages = {153--164},
  year = {2016},
  address = {San Jose, CA, USA},
  publisher = {ACM},
  location = {New York}
}

@article{gelman2011concepts,
  author = {Gelman, Susan A. and Legare, Cristine H.},
  title = {Concepts and folk theories},
  journal = {Annual Review of Anthropology},
  volume = {40},
  pages = {379--398},
  year = {2011}
}

@book{gray2019ghost,
  author = {Gray, Mary L. and Suri, Siddharth},
  title = {Ghost Work: How to Stop Silicon Valley from Building a New Global Underclass},
  publisher = {Houghton Mifflin Harcourt},
  address = {Boston},
  year = {2019}
}

@article{guest2006how,
  author = {Guest, Greg and Bunce, Arwen and Johnson, Laura},
  title = {How many interviews are enough? An experiment with data saturation and variability},
  journal = {Field Methods},
  volume = {18},
  number = {1},
  pages = {59--82},
  year = {2006}
}

@inproceedings{katell2020toward,
  author = {Katell, Michael and Young, Meg and Dailey, Dharma and Herman, Bernease and Guetber, Vivian and Tam, Aaron and Binz, Corinne and Raz, Daniella and Krafft, P. M.},
  title = {Toward situated interventions for algorithmic equity: Lessons from the field},
  booktitle = {Proceedings of the 2020 Conference on Fairness, Accountability, and Transparency},
  pages = {45--55},
  year = {2020},
  address = {Barcelona, Spain},
  publisher = {ACM},
  location = {New York}
}

@book{pasquale2015black,
  author = {Pasquale, Frank},
  title = {The Black Box Society: The Secret Algorithms That Control Money and Information},
  publisher = {Harvard University Press},
  address = {Cambridge, MA},
  year = {2015}
}

@inproceedings{rader2015understanding,
  author = {Rader, Emilee and Gray, Rebecca},
  title = {Understanding user beliefs about algorithmic curation in the Facebook news feed},
  booktitle = {Proceedings of the 33rd Annual ACM Conference on Human Factors in Computing Systems},
  pages = {173--182},
  year = {2015},
  address = {Seoul, Republic of Korea},
  publisher = {ACM},
  location = {New York}
}

@article{siles2020folk,
  author = {Siles, Ignacio and Segura-Castillo, Andr{\'e}s and Sol{\'i}s, Ricardo and Sancho, M{\'o}nica},
  title = {Folk theories of algorithmic recommendations on Spotify: Enacting data assemblages in the global South},
  journal = {Big Data \& Society},
  volume = {7},
  number = {1},
  pages = {1--15},
  year = {2020}
}

@inproceedings{sloane2022participation,
  author = {Sloane, Mona and Moss, Emanuel and Awomolo, Olaitan and Forlano, Laura},
  title = {Participation is not a design fix for machine learning},
  booktitle = {Proceedings of the 2022 AAAI/ACM Conference on AI, Ethics, and Society},
  pages = {702--709},
  year = {2022},
  address = {Oxford, UK},
  publisher = {ACM},
  location = {New York}
}

@article{clarke2003situational,
  title={Situational analyses: Grounded theory mapping after the postmodern turn},
  author={Clarke, Adele E},
  journal={Symbolic interaction},
  volume={26},
  number={4},
  pages={553--576},
  year={2003},
  publisher={Wiley Online Library}
}

@article{braun2006using,
  title={Using thematic analysis in psychology},
  author={Braun, Virginia and Clarke, Victoria},
  journal={Qualitative research in psychology},
  volume={3},
  number={2},
  pages={77--101},
  year={2006},
  publisher={Taylor \& Francis}
}

@book{barad2007meeting,
  title={Meeting the universe halfway: Quantum physics and the entanglement of matter and meaning},
  author={Barad, Karen},
  year={2007},
  publisher={duke university Press}
}

@book{kennedy2016post,
  title={Post, mine, repeat: Social media data mining becomes ordinary},
  author={Kennedy, Helen},
  year={2016},
  publisher={Springer}
}

@article{pinch1984social,
  title={The social construction of facts and artefacts: Or how the sociology of science and the sociology of technology might benefit each other},
  author={Pinch, Trevor J and Bijker, Wiebe E},
  journal={Social studies of science},
  volume={14},
  number={3},
  pages={399--441},
  year={1984},
  publisher={Sage Publications}
}

@incollection{couldry2019costs,
  title={The costs of connection: How data is colonizing human life and appropriating it for capitalism},
  author={Couldry, Nick and Mejias, Ulises A},
  booktitle={The costs of connection},
  year={2019},
  publisher={Stanford University Press}
}

@article{reilhac2017user,
  title={User experience with increasing levels of vehicle automation: Overview of the challenges and opportunities as vehicles progress from partial to high automation},
  author={Reilhac, Patrice and Hottelart, Katharina and Diederichs, Frederik and Nowakowski, Christopher},
  journal={Automotive User Interfaces: Creating Interactive Experiences in the Car},
  pages={457--482},
  year={2017},
  publisher={Springer}
}

@article{hind_dashboard_2021,
	title = {Dashboard design and the ‘datafied’ driving experience},
	volume = {8},
	issn = {2053-9517},
	url = {https://doi.org/10.1177/20539517211049862},
	doi = {10.1177/20539517211049862},
	language = {EN},
	number = {2},
	urldate = {2026-01-30},
	journal = {Big Data \& Society},
	author = {Hind, Sam},
	month = jul,
	year = {2021},
	note = {Publisher: SAGE Publications Ltd},
	pages = {20539517211049862},
}

@article{crawford2016can,
  title={Can an algorithm be agonistic? Ten scenes from life in calculated publics},
  author={Crawford, Kate},
  journal={Science, Technology, \& Human Values},
  volume={41},
  number={1},
  pages={77--92},
  year={2016},
  publisher={SAGE Publications Sage CA: Los Angeles, CA}
}

@book{eubanks2018automating,
  title={Automating inequality: How high-tech tools profile, police, and punish the poor},
  author={Eubanks, Virginia},
  year={2018},
  publisher={Macmillan+ ORM}
}

@article{beraldo2019data,
  title={From data politics to the contentious politics of data},
  author={Beraldo, Davide and Milan, Stefania},
  journal={Big Data \& Society},
  volume={6},
  number={2},
  pages={2053951719885967},
  year={2019},
  publisher={SAGE Publications Sage UK: London, England}
}

@article{stilgoe2018machine,
  title={Machine learning, social learning and the governance of self-driving cars},
  author={Stilgoe, Jack},
  journal={Social studies of science},
  volume={48},
  number={1},
  pages={25--56},
  year={2018},
  publisher={SAGE Publications Sage UK: London, England}
}

@article{bradshaw2013seven,
  title={The seven deadly myths of" autonomous systems"},
  author={Bradshaw, Jeffrey M and Hoffman, Robert R and Woods, David D and Johnson, Matthew},
  journal={IEEE Intelligent Systems},
  volume={28},
  number={3},
  pages={54--61},
  year={2013},
  publisher={IEEE}
}

@book{winner1978autonomous,
  title={Autonomous technology: Technics-out-of-control as a theme in political thought},
  author={Winner, Langdon},
  year={1978},
  publisher={Mit Press}
}

@book{saldana2011fundamentals,
  title={Fundamentals of qualitative research},
  author={Saldana, Johnny},
  year={2011},
  publisher={Oxford university press}
}

@article{lupton2016digital,
  title={Digital companion species and eating data: Implications for theorising digital data--human assemblages},
  author={Lupton, Deborah},
  journal={Big Data \& Society},
  volume={3},
  number={1},
  pages={2053951715619947},
  year={2016},
  publisher={SAGE Publications Sage UK: London, England}
}

@article{schulz2025generative,
  title={Why generative AI is different from designed technology regarding task-relatedness, user interaction, and agency},
  author={Schulz-Schaeffer, Ingo},
  journal={Big Data \& Society},
  volume={12},
  number={3},
  pages={20539517251367452},
  year={2025},
  publisher={SAGE Publications Sage UK: London, England}
}

@article{zhao2025boosting,
  title={Boosting popularity: Folk theories and algorithmic resistance of visibility contests in the comment sections},
  author={Zhao, Guoning},
  journal={Big Data \& Society},
  volume={12},
  number={2},
  pages={20539517251331949},
  year={2025},
  publisher={SAGE Publications Sage UK: London, England}
}

@article{couldry2019data,
  title={Data colonialism: Rethinking big data’s relation to the contemporary subject},
  author={Couldry, Nick and Mejias, Ulises A},
  journal={Television \& new media},
  volume={20},
  number={4},
  pages={336--349},
  year={2019},
  publisher={Sage Publications Sage CA: Los Angeles, CA}
}

@incollection{elish2025moral,
  title={Moral crumple zones: cautionary tales in human--robot interaction},
  author={Elish, Madeleine Clare},
  booktitle={Robot Law: Volume II},
  pages={83--105},
  year={2025},
  publisher={Edward Elgar Publishing}
}

@article{plantin2018re,
  title={Re-integrating scholarly infrastructure: The ambiguous role of data sharing platforms},
  author={Plantin, Jean-Christophe and Lagoze, Carl and Edwards, Paul N},
  journal={Big Data \& Society},
  volume={5},
  number={1},
  pages={2053951718756683},
  year={2018},
  publisher={SAGE Publications Sage UK: London, England}
}

@book{book,
  editor = {Lie, Merete and S{\o}rensen, Knut H.},
  title = {Making Technology Our Own? Domesticating Technology into Everyday Life},
  year = {1996},
  publisher = {Scandinavian University Press},
  address = {Oslo}
}

@book{haraway1991simians,
  title={Simians, Cyborgs, and Women: The Reinvention of Nature},
  author={Haraway, Donna J.},
  year={1991},
  publisher={Routledge},
  address={New York},
  isbn={9780415903875}
}

@article{bates2025feeding,
  title={Feeding the machine: Practitioner experiences of efforts to overcome AI's data dilemma},
  author={Bates, Jo and Fratczak, Monika and Kennedy, Helen and Perea, Itzelle Medina and Ochu, Erinma},
  journal={Big Data \& Society},
  volume={12},
  number={4},
  pages={20539517251396092},
  year={2025},
  publisher={SAGE Publications Sage UK: London, England}
}

\end{document}